\documentclass[12pt]{article}
\usepackage{amsfonts}
\usepackage{eurosym}
\usepackage{graphicx}
\usepackage{booktabs}
\usepackage{float}
\usepackage{subfig}
\usepackage{amssymb}
\usepackage{comment}
\usepackage{amsmath}
\usepackage{booktabs}
\usepackage{multirow}
\usepackage{enumerate}
\usepackage{color}
\usepackage[authoryear]{natbib}
\usepackage{hyperref}

\hypersetup{colorlinks=true, linkcolor=red, citecolor=blue}

\begin{document}

\title{A Typology of Social Capital and Associated Network Measures\thanks{This was written for a special issue in memory of Kenneth J. Arrow.  Conversations with Ken about social capital
and the role of networks helped sharpen my thinking on the subject.
It is sad not to have the opportunity to discuss this paper with him.
}}
\author{Matthew O. Jackson \thanks{%
Department of
Economics, Stanford University, Stanford, California 94305-6072 USA,
external faculty member at the Santa Fe Institute, and a fellow of CIFAR.
Email:  and jacksonm@stanford.edu.
I gratefully
acknowledge financial support under ARO MURI Award No. W911NF-12-1-0509 and NSF grant SES-1629446.
I also thank Robert Fluegge,  Eduardo Laguna-Muggenberg, Mihai Manea, Sharon Shiao, and two anonymous referees for helpful comments on earlier drafts.}}
\date{Draft: February 2019}
\maketitle

\begin{abstract}

I provide a typology of social capital, breaking it down into seven more fundamental forms of capital:  information capital, brokerage capital, coordination and leadership capital, bridging capital, favor capital, reputation capital, and community capital.
I discuss how most of these forms of social capital can be identified using different network-based measures.


\end{abstract}

\thispagestyle{empty}

\setcounter{page}{0} \newpage

\section{Introduction}

Social capital plays a central role
 in a multitude of production processes, both formal and informal, from the founding of a company, to the reaching out to get help
 in a product design,  to the negotiations and compromise that produces political legislation, to how workers get hired, to whether people educate themselves,.
 It is fundamental to the welfare of a society, and its distribution is important in driving inequality and immobility.\footnote{For instance, see the discussion in \cite{knackk1997,jackson2018}.}
Despite the importance of social capital in almost every production process, it remains a remarkably murky concept.

Part of the murkiness comes from the many different ways in which social capital has been defined over the last century.
An early explicit mention of social capital is by Lydia J. Hanifan \citeyearpar{hanifan1916,hanifan1920} who wrote:\footnote{One can find earlier mentions of the term - going back to the late 1800s - but often with different meanings from the more modern ones. }
``In the use of the phrase social {\sl capital} I make no reference to the usual acceptation of the term {\sl capital}, except in a figurative sense. I do not refer to real estate, or to personal property or to cold cash, but rather to that in life which tends to make these tangible substances count for most in the daily lives of a people, namely, goodwill, fellowship, mutual sympathy and social intercourse among a group of individuals and families who make up a social unit...''
Hanifan's definition is broad and does not make it completely obvious why this would be a form of capital.
Two later, and widely-cited, definitions of  social capital, by Glenn Loury and Pierre Bourdieu, take different angles from Hanifan, and from each other; but make it clearer that some form of capital is involved.
Glenn Loury states \citep{loury1977} ``It may
thus be useful to employ a concept of ``social capital'' to represent the
consequences of social position in facilitating acquisition of the standard human
capital characteristics.''   Loury's definition is much more precise and it becomes clearer why this would be a form a capital, as it helps in the production of education.  But the definition restricts the use of social capital to acquiring human capital, which excludes many other applications in which social capital plays a central role.
Pierre Bourdieu defined it as \citep{bourdieu1986} 
``the aggregate of the actual or potential
resources which are linked to possession of a durable network of more
or less institutionalized relationships of mutual acquaintance or recognition...''
This admits more settings than Loury's definition as it applies to general resources, but more specifically founds the definition of social capital in networks of relationships.  Despite admitting
a broader set of applications, Bourdieu's definition is still sufficiently vague that it is hard to know what it really is or how to measure it in practice.

These are just a few of the enormous number of definitions,\footnote{  See, for instance, the many definitions in \cite*{portes1998,woolcock1998,dasguptas2001,sobel2002,glaeserls2002,dasgupta2005}.  That important forms of capital are embodied in humans as well as their relationships is
even mentioned in the writings of Alfred Marshall.} but illustrate how varied and difficult to work with the definitions can be.

The purpose of the current paper is to offer a typology of forms of capital that all fit under a broad umbrella of social capital.  By making
more precise the variety of different things that can all be considered social capital, we can clarify the broader concept and also provide a foundation for working with specific forms of social capital.
Beyond verbal definitions, I also provide associated network measures.  These measures help operationalize the definitions to help eliminate ambiguity.

Before previewing the definitions and typology of social capital, it is useful to discuss what it means to be `capital', as this helps in narrowing down the definitions an understanding why many definitions of social capital are ambiguous.

The importance of social capital and the difficulties with its
 definition did not escape Ken Arrow, as almost nothing did.      
Some of his thoughts about social capital appear in \cite{arrow2000},
in a volume collected around a conference on social capital definitions
 and applications.  Ken (as well as Bob \cite{solow2000}) made the point that the analogy to ``capital'' is a stretch, and Ken even suggests abandoning the analogy.
Ken states (page 4):
 ``The term ``capital'' implies three aspects:
(a) extension in time; (b) deliberate sacrifice in the present for
future benefit; and (c) alienability.''
Ken went on to note that (c) fails for social capital, but also for human capital and which he implicitly does not want to abandon as a form of capital.
It is also not at all obvious why alienability should be a prerequisite to call something capital (and more on this below).
Ken goes on to state: ``But it is especially (b) that fails. The essence of social networks is that
they are built up for reasons other than their economic value to the participants.''

Before digging more deeply into what it means to be a form of capital, I take exception with Ken's discussion of ``(b) deliberate current sacrifice for future benefits.''
First, just as one example, people can be very judicious in asking friends for favors (a form of social capital discussed below) as they realize that potential favors are scarce and may be of greater value for some future need.
Second,
there are many settings in which people deliberately build relationships for economic value and purposes: the term ``networking'' in its business connotation refers to this explicitly.  Part of the perceived value of a Harvard Business School education - something that many people are willing to pay dearly for - is the set of relationships that are built with classmates.
Third, it is not clear why decisions to accumulate, store, and use something have to be `deliberate' in order to consider something capital.
Although relationships may be built for a variety of reasons that do not include deliberate consideration of the information or other benefits that may eventually accrue from them,
this is also true of many other forms of capital.  If I planted a tree in order to consume its beauty, but then years later found that its bark could be used in producing a drug to cure cancer, that tree would still count as capital
despite the fact that this was not the intention when planting of the tree.

These points make it clear that it will be useful to review the origins of the general term ``capital'' and to understand how it relates to production processes,  and I start by examining  Adam Smith's \citeyearpar{smith1776} definition.

Adam Smith   starts by discussing stocks of goods (page 212):  ``But when the division of labour has once been thoroughly introduced, the produce of a man's own labour can supply but a very small part of his occasional wants.  The far greater part of them are supplied by the produce of other men's labour, which he purchases with the produce, or , what is the same thing, with the price of the produce of his own.  But this purchase cannot be made till such time as the produce of his own labour has not only been completed, but sold.   A stock of goods of different kinds, therefore, must be stored up somewhere sufficient to maintain him, and to supply him with the materials and tools of his work till such time, at least, as both these events can be brought about.  ''
Smith then goes on to define capital (page 214):   ``But when he possesses stock sufficient to maintain him for months or years, he naturally endeavours to derive a revenue from the greater part of it; reserving only so much for his immediate consumption as may maintain him till this revenue begins to come in.  His whole stock, therefore is distinguished into two parts.  That part which, he expects, is to afford him this revenue, is called his capital.''

Smith's notion of capital were goods that are produced and can be stored and used in further production.  For him this included a variety of things such as grains, livestock, plants, wood, furniture, tools, and machines.
Deliberately excluded were land and labor.

Although Smith did discuss importance of skill, he did not fully anticipate
human capital, and his writings are strained by the fact that not all labor is the same.
Alfred Marshall later discussed personal capital - essentially human capital - recognizing that skills and knowledge were a stock that took investments to produce and could yield returns and were valuable inputs into production separate from the pure efforts of labor.  However, he wavered on the topic and eventually took the definition of personal capital out of his Principles.\footnote{Marshall's views were complex and changed over time, e.g., see the discussion of \cite{blandy1967}.}

To see how to fit things like human capital and social capital into a definition of capital, it is useful to state a modern definition of capital, and to state it succinctly and directly.
Here I define `capital' to be {\sl any} stock - other than land and labor - that can be used, or converted into something that is useful in the production or distribution of {\sl any} good, service, skill, or knowledge.\footnote{The exclusion of land and labor as forms of capital is mainly made to stick with the historical distinction.  That historical distinction was made mostly for convenience - it can be very useful to distinguish capital and labor in estimating production functions and in many
modeling and policy applications.  Given that land and labor play no role in the discussion here, I exclude them for ease of exposition and to stick with the classical roots of the definition, but not for some deeper logical reason.}$^,$\footnote{Here I also separate distribution from production, just so that people understand that it is included, although one could fold it into the definition of production if that term is suitably understood.}

Under this definition of `capital', many forms of social capital fit in.   For instance, friendships that can be counted on to provide favors and knowledge
 are stocks in the sense that they exist in measurable quantities that last over time,
and they can called on for productive purposes.  Moreover, they have other attributes that classical forms of capital have.
The person who `possesses' those fruitful friends can choose whether to consume favors and knowledge now, or to wait until later.  The relationships can also depreciate with time.
For instance, when examining `favor capital' (defined in more detail below), a person who has built friendships and provided help to friends in the past can call on those friends for favors.  Those favors can be used
for productive purposes: I may call on a friend to lend me tools for a construction project, or for a loan, or to provide feedback and advice on a research project.  The favor capital is essentially a claim that I have on some sort of other physical, financial, or human capital - so it is as if I own at least some part of it.
It is not in endless supply - I will lose the friendships if I ask for too many favors without reciprocation. It depreciates:  if I have not put any efforts into a friendship over some period of time, the friend may not be as willing to grant a favor when asked.
This, again, counters Ken's criticism that social capital is not something for which there is deliberate current sacrifice for future benefits.

It is clear from their discussions that part of Arrow and Solow's frustration with the term `capital' applied to social settings stems from the imprecision of definitions of social capital (see also \cite{durlauf1999,sobel2002} on this issue), which make it hard to see what social capital actually is, much less why it would be capital.
Indeed, despite the ubiquity of the concept and its applications, its definitions are often confusing and end up mixing different concepts.
In some cases they lean on other terms like ``goodwill'' or ``the value that is embedded in social relationships'' that can be vague out of context.
But the main challenge in coming up with a useful definition of social capital is that there are many types of social interactions and they play different roles - and thus there are many different types of social capital, and without properly distinguishing them it is hard to know what is being referred to.

The contribution here is not in discovering new forms of social capital that have never been seen before, but rather in providing a logical structure and order to the umbrella concept of social capital by being explicit about its different types and how they can be measured.
There is a balance between being useful and exhaustive; therefore, I have tried to distinguish types of capital that are clearly different, but without distinguishing every nuance.
This results in seven different concepts.

As a brief preview (see below for additional discussion and references), the seven different types of social capital that I distinguish are:
\begin{itemize}
\item  {\bf Information Capital}:  the ability to acquire valuable information and/or to spread it to other people who can use it through social connections.
\item  {\bf Brokerage Capital}:  being in a position to serve as an intermediary between others who wish to interact or transact.
\item  {\bf Coordination and Leadership Capital}:  being connected to others who do not interact with each other, and having the ability to coordinate others' behaviors.
\item  {\bf Bridging Capital}:  being an exclusive connector between otherwise disparate groups, with an ability to acquire, as well as control the flow of, valuable
information.
\item  {\bf Favor Capital}:   the ability to exchange favors and safely transact with
others through a combination of network position and repeated interaction and reciprocation.
\item {\bf Reputation Capital}:
having others believe that a person or organization is reliable and/or provides consistently high quality advice, information, labor, goods, or services.
\item  {\bf Community Capital}:  the ability to sustain cooperative (aggregate social-welfare-maximizing) behavior in the running of institutions, the provision of public goods, the handling of commons, and/or collective action, within a community.
\end{itemize}

Let me emphasize that something like ``information capital'' is not meant to be thought of as some form of capital that is only used in producing information. Instead it should be viewed as the access to information that is used in {\sl any} production process, including quite standard ones.\footnote{This is also another source of confusion in the use of the term `social capital', as it is often invoked in contexts relative to the improvement of an individual, which is certainly part of the story, but without reference to any production.  Ultimately we care about how capital is instrumental in the  ``production'' of something - a good, service, entertainment, or anything else that can be consumed or enjoyed by someone.   Thus, something like favor capital can be possessed by an individual, but the value comes from the productive uses that it can eventually enable.  } 
For instance, if some software engineer is producing computer code and needs to figure out how to structure some database and does not personally know how to do it,   the ``information capital'' that the engineer has would refer to the contacts and resources that the engineer can call upon to help him or her structure that database and ultimately produce the software.   That form of capital is just as vital to production as the engineer's human capital (stored personal knowledge) and physical capital (the computer that he or she uses to code on,  space they occupy, chair they sit on, etc.).  
Thus, rather than simply referring to the engineer's ``social capital'' being important in helping them produce the code, ``information capital'' makes it more precise that it is access to information and knowledge that is the input, rather than some other form of favor or social help that might be used in production.

Part of the confusion around definitions of social capital also stems from the fact that there is a wide variety of forms of ``production'' to which it applies,
and it is often applied  to discuss whether some individual or group will be ``successful'' and not in the explicit context of any production.    Production comes in many forms.   For instance, an individual may want to ``produce'' a job for herself - and may draw on  information and/or favors to obtain a job.
Thus, whether or not she is able to produce a job for herself depends on her access to information and favor capital.    This particular is outside of the realm of a classic good or service, but can be viewed as valuable production for the individual.\footnote{Note that some uses could also have negative value, for instance some forms of favor capital might include nepotism that has negative externalities.  But many forms of production involve externalities, and that does not change the discussion of whether something is an essential and valued input to that production process. }  It makes clear how broad the set of production processes that can be considered are.

The network-based measures that I define below to capture these concepts either have cousins in the centrality literature, or are adjustments or applications of existing centrality measures.\footnote{Network measures of centrality, influence, and power are abundant  (e.g., see
\cite*{borgatti2005,jackson2008,blochjt2016,jackson2018}).  Although there is much discussion of centrality measures in the literature, the explicit discussion of how different ones may be used to identify different forms of social capital is new to this paper.}
I have made these adjustments with two things in mind:  matching measures to the concepts of capital that they are intended to capture, which can help clarify the precise meaning; and providing
measures that are computable in the increasingly large data sets the accompany many applications.
Also, not all seven forms of social capital are associated with networks of relationships, and so the network measures that I provide are for the first five on the list, and the last two require other sorts of observations.

I will not try to make sense of the full literature on social capital here -- as it is sprawling and inconsistent in its multitude of
alternative definitions  and uses of the term.  I will mention references as needed.
The motivation of the current paper is to be precise in defining particular types of social capital, and then viewing social capital as an umbrella concept. This will help avoid the confusion that stems from people's use of the loose general term even though they have different narrower types of it in mind.

Nonetheless, it is important to recognize that the literature has made distinctions between whether social capital applies to an individual or a community (e.g., see \cite*{borgattije1998}).  This difference can already be seen in the contrast between Hanifan's definition above in which social capital is something possessed by a community and Loury's definition which is more distinctly about a given individual's position.  These are very different things.
The main definitions that I provide below are forms of capital that accrue to an individual.  They can be aggregated to a community level, but are distinct from the idea of a community that functions well.
That sort of community capital is also important, and is very distinct from these other forms of social capital.  I discuss community-based social capital and the distinction in Section \ref{community}.

I am also certainly not the first to recognize that there are different forms of social capital.
For example, one can find distinctions between whether social capital is involved in `bridging' or `bonding' (\cite{szreterw2004}).\footnote{See also \cite*{lin1999,flapv2001,kawachietal2004,aldrich2012}.}
Again, the contribution here is to provide a careful typology together with explicit definitions and measures for each concept.

\section{Background Definitions and Notation}

I begin by providing some notation that will help in defining measures.

There are $n$ individuals indexed by $i\in \{1,2,..n\}$.  Depending on the context these `nodes' may be individual people, or they may be a group, or an organization.

A \emph{network} is a graph, represented by its adjacency matrix $\mathbf{g}\in [0,1]^{n\times n}$,
where $g_{ij}>0$ indicates the existence of an edge (a.k.a. link, tie, connection...) between $i$ and $%
j$ and $g_{ij}=0$ indicates the absence of a edge.\footnote{Generally, I will consider a case in which $g_{ii}=0$, so that there are no self-loops, but this is not consequential to the formal definitions.}

In most of what follows I will discuss definitions for the case of simple graphs: undirected and un-weighted networks.
Nonetheless, the definitions cover the weighted and/or directed cases as well.
When modifications are necessary, I will indicate those details.

Let $G(n)$ denote the set of all admissible networks on $n$ nodes.

The \emph{degree} of a node $i$ in a network $\mathbf{g}$, denoted $d_i(\mathbf{g})=\sum_{j} g_{ij}$, is the number of edges
involving node $i$.  In the case of a directed network, this is $i$'s out-degree and indegree is $\sum_{j} g_{ji}$.

A \emph{walk} between $i$ and $j$ is a
succession of (not necessarily distinct) nodes $i=i^0, i^1,...,i^M=j$ such
that $g_{i^m i^{m+1}} = 1$ for all $m=0,..,M-1$.

A \emph{path} in $\mathbf{g}$ between two nodes $i$ and $j$  is a
succession of distinct nodes $i=i^0, i^1,...,i^M=j$ such that $%
g_{i^mi^{m+1}} = 1$ for all $m=0,..,M-1$.

Two nodes $i$ and $j$ are
connected (or path-connected) if there exists a path between them.

A \emph{geodesic} (shortest path)
between nodes $i$ and $j$ is a path such that no other path between them involves a
smaller number of edges.

The number of geodesics between $i$ and $j$ is denoted $%
\nu_{\mathbf{g}}(i,j)$.  Let $\nu_{\mathbf{g}}(k:i,j)$ denote the number of geodesics between $%
i $ and $j$ passing through $k$.

The \emph{distance} between nodes $i$ and $j$, denoted $%
\ell_{\mathbf{g}}(i,j)$, is the number of edges involved in a geodesic  between $i$
and $j$.
This  is defined only for pairs of nodes that have a path between them and may be taken to be $\infty$ otherwise.

Note that $[\mathbf{g}^\ell]_{ij}$ counts the number of walks of length $\ell$
between nodes $i$ and $j$.

Let $N^{\ell}_i(\mathbf{g})$ be the set of individuals at distance $\ell$ from $i$ in network $g$:   $ N_i^\ell(\mathbf{g})= \{j: \ell(i,j)=\ell \}$.

Let $N_i(\mathbf{g})= N_i^1(\mathbf{g})$.

Node $i$'s degree is $d_i(\mathbf{g}) = |N_i(\mathbf{g})|$.

Similarly, let $n^\ell_i(\mathbf{g})=|N_i^\ell(\mathbf{g})|$ denote a higher order degree:  the number
of nodes at distance $\ell$ from node $i$.

Let $clust_i(\mathbf{g})$ denote the clustering of node $i$:   the fraction of pairs of $i$'s neighbors who are connected to each other:
$\sum_{kj\in N_i(\mathbf{g}), k<j } \frac{g_{kj}}{d_i(\mathbf{g})(d_i(\mathbf{g})-1)/2}$

\section{Social Capital Definitions and the Network Measures}

I now discuss the different concepts of social capital and various measures of the concepts.

\subsection{Information Capital}

One of the most important roles that we obtain from our friends and acquaintances is information.   From information about jobs to tips on raising children to details of how to better provide a good or service, much of the information has significant productive value. \footnote{For example,
\cite{arrowb2004} discuss how having more access to job information can result in more productive matching between workers and jobs.}
In reverse, it can also be very productive to be able to spread information - from information about products and programs to information about available jobs.
Of course, acquiring and spreading are two different types of information capital.
Both are important aspects of production: being able to gather knowledge that can improve productivity, as well as getting that information into other's hands.
In undirected networks, these will have very similar signatures and will be captured by the same measures; but in directed networks these have reversed measures.

The basic definition of information capital --  an individual's ability to acquire valuable information as well as to spread it to other people who can use it through social connections -- includes both directions.
The associated measures of information capital based on networks account for how many people an individual can either send information to or receive it from.

The key idea here is that the chance that information is relayed decays with social distance.    This reflects two things.
The first is that interaction is stochastic, so there is only a chance that two people who are friends exchange a given piece of information.
The second is that the information may degrade and become less useful as it is repeatedly passed (as we all know from playing the classic game of `telephone').\footnote{For another reason for such a decay, see \cite{manea2017} who examines a model of resale of information with bargaining.}

The decay of information with distance is captured via a parameter $p$, where we will generally presume that $0<p<1$.   The concepts below are still well-defined without this assumption.

In addition, we will also include a parameter $T$ that caps the number of times that information is relayed - we may think of this
as the information's `endurance'.   This may reflect information becoming stale.

As these parameters vary, the calculation of information capital will vary.  Thus, the measure for high $p$ and/or $T$ can look very different from a setting with lower parameters.    This means that an individual may have a lot of information capital when it comes to long lasting and high $p$ topics, but have low social capital for topics with substantial decay and low endurance.

The simplest version of a network centrality index that can be used to measure information capital is that of
decay centrality (defined by \cite{jackson2008} and related to the connections model of \cite{jacksonw1996}):
\[
Dec_i(\mathbf{g},p,T) = \sum_{\ell=1}^Tp^{\ell}
|N^{\ell}_i(\mathbf{g})| .
\]

Decay centrality counts paths of different lengths - so how many people one can reach at different distances - and weights them by their distance.  Someone at distance 2 counts $p^2$, which can be much less than someone at distance 1 who counts $p$.
One way to think to interpret the measure is that one can easily ask one's friends for information.  It becomes harder to make contact with friends of friends and get information from them.   The relative additional difficulty in reaching out to a friend of a friend is governed by $p$.
Friends of friends of friends are even more difficult to reach out to.
At some point, it is no longer feasible to reach out to a person if they are at too large a distance, and hence the cap $T$.

Thus, decay centrality might be thought of as a measure of how many people someone could reach out to for helpful information, appropriately weighted for the difficulty in reaching them.
Decay centrality also works in reverse, as in broadcasting information, as it captures the ease of getting information to someone, by similar reasoning.
People talk about topics, and information percolates through the network.  Whether one is effective at either reaching many people, or hearing from many people, depends on one's network position and how many hops it takes to reach various other people.

An advantage of decay centrality is that it is easy to measure.
A disadvantage of decay centrality is that it does not account for the fact that in some settings it is easier to reach any other given person, either in sending or receiving information,
if there are many independent paths to that person.
Multiple paths can increase the chance that the information makes it from one node to another.   This was the motivation behind the concept of communication centrality, as defined by \cite*{banerjeecdj2013}.\footnote{See also the related notion of cascade centrality from \cite*{kempekt2003,kempekt2005}.}

Communication centrality is defined as follows.

It is based on a simple diffusion process.    Some information starts at node $i$.   Node $i$ passes this information to a node $j$ with probability $p_{ij}\in [0,1]$ in a given time period.    So, $\mathbf{p}$ is an $n\times n$ weighted and directed network with $ij$th entry being the probability that $i$ passes information to $j$ in a given time period.

A special case is in which $p_{ij}$'s are identical for all $ij$ and then  $\mathbf{p}$ becomes $p \mathbf{g}$, where $p$ is a scalar and $\mathbf{g}$ is an adjacency matrix taking on values $\{0,1\}$.

Define $PInf(\mathbf{p},T)_{ij}$ to be the probability that node $j$ ends up hearing information that starts from node $i$ if it is passed independently with probability $\mathbf{p}_{i'j'}$ from node $i'$ to $j'$ along each walk in the network, and running the whole process for $T$ periods.

Then communication centrality is
\[
Com_i(\mathbf{p},T) = \sum_j  PInf(\mathbf{p},T)_{ij}.
\]

Communication centrality involves some simulation to calculate, as one needs to account for all the possible paths that information might take, and some end up overlapping, producing correlation in the chance that information makes it from one node to another.
A measure that is richer than decay centrality in accounting for multiple paths, but easier to work with than communication centrality is
diffusion centrality.  It is defined as follows.

First, let
\[
EInf(\mathbf{p},T)_{ij} = \sum_{\ell=1}^T  \left[\mathbf{p}^\ell\right]_{ij}.
\]
This is the expected number of times that $j$ will hear information that starts at node $i$ and is passed according to the matrix $\mathbf{p}$ for $T$ periods.
$EInf$ differs from $Pinf$ by counting multiple hearings, and so can be greater than 1, while $PInf$ is just a probability of ever hearing.
For instance, if information makes it from $i$ to $j$ via two different routes, then both are counted, and so rather just accounting for the fact that information might make it from $i$ to $j$, the measure gives extra points if the information makes it several times.

$EInf$ is `relatively' easy to calculate as it only involves raising a matrix to a power, and in cases where multiple hearings of information are more valuable or influential than hearing just once, then it can even be more appropriate.

Diffusion centrality  (from \cite*{banerjeecdj2013}) is then defined as:
\[
Diff_i(\mathbf{p},T) = \sum_j  EInf(\mathbf{p},T)_{ij}  =  \sum_j
 \sum_{\ell=1}^T   [\mathbf{p}^\ell]_{ij}.
\]

For low levels of $\mathbf{p}$ and/or $T$,  decay, communication and diffusion centrality can be very similar, but then they start to diverge when the values of $\mathbf{p}$ and $T$ increase.\footnote{Communication and diffusion centrality allow for heterogeneous weightings, while decay centrality does not.  However,
it similarly extends by tracking shortest weighted paths.}
In the extreme, when $T=1$ so that information does not travel beyond immediate acquaintances, and $\mathbf{p}$ takes on values $\{0,p\}$ for some scalar $p$ so that
 people pass information with the same probability, then all three measures reduce to
a node's (out) degree.

As $T$ grows,  diffusion centrality converges to either eigenvector centrality (if $p$ is at least as large as the inverse of the first
eigenvalue of $\mathbf{g}$) or Bonacich centrality (if $p$ is smaller than the inverse of the first
eigenvalue of $\mathbf{g}$), as proven by \cite*{banerjeecdj2015}.

\subsubsection{Valued Relationships}

The centrality measures above focus on the movement of information.
In some settings, we also need to account for the stock of information that is held by individuals in a society, as well as how useful that information can be to the receivers.\footnote{For instance, see \cite{atkesonk1993}  who consider the impact of the knowledge embodied in a group or organization.  }

In particular, in many settings there are different values for the information that comes from different nodes,
or in getting information to different nodes.    For instance, if one wants medical information, then it may only be paths to medical professionals that generate information.
  Similarly, in spreading information about a government program, that information is more valuable to people who are eligible for the program.  Thus, one can adjust these measures to have weightings that reflect the variation in the value of accessing different nodes.\footnote{See \cite{katz1953,bonacich1987,jackson2008,laguna2017} for discussion.   \cite{laguna2017} shows that including values can make a big difference in who is most central, especially in networks with homophily.}

Let $\mathbf{v}$ be the $n\times n$ matrix in which entry $v_{ij}$ is the value that $j$ gets of information that comes from $i$.  Of course, this matrix $\mathbf{v}$ depends on circumstances.  If a given receiving node is healthy, then information from a doctor might not be valuable, while if the receiving node is ill then the information is more valuable.
A default is just to think of all $v_{ij}$'s to be the same, so that $\mathbf{v}$ drops out of the calculations.
But it might be that some forms of cross information are more valuable than others -- and this becomes especially important when we think of the potential
synergies that information might bring.   Computational methods could be very valuable to a linguist, and not something easily obtained from other linguists.  In that case, if $j$ is a linguist then $v_{ij}$ would be high if $i$ had the knowledge of the methods needed by $j$, but might be very low if $i$ does not have that knowledge.

The respective generalizations are
\[
Dec_i(\mathbf{g},p,T, \mathbf{v}) =  \sum_{\ell=1}^T  p^{\ell} \sum_{j: j \in N^\ell_i (\mathbf{g})}
 v_{ij},
\]
\[
Com_i(\mathbf{p},T, \mathbf{v}) = \sum_j  PInf(\mathbf{p},T)_{ij} v_{ij},
\]
\[
Diff_i(\mathbf{p},T, \mathbf{v}) = \sum_{\ell=1}^T  \sum_j
[\mathbf{p}^\ell]_{ij}v_{ij}.
\]

\subsubsection{Large Society Approximations}

When faced with large networks, the calculations, especially for large $T$, become computationally taxing.    In such cases, truncating the calculation at
some lower $T'$ - for instance 2 or 3, so that one only has to do local neighborhood expansions - can provide an approximation.
Such an approximation is obviously more accurate for lower levels of $\mathbf{p}$, in which case closer neighborhoods have substantially higher weight and longer paths and walks become more heavily discounted.\footnote{This requires that the entries of $\mathbf{p}$ be substantially lower than one over the average degree,  so that closer neighborhoods are accounting for relatively more of the calculations.  More generally  .}
For instance, counting friends and friends of friends, but then ignoring neighborhoods beyond the second degree is one approach.

Alternatively, one can also estimate further expansions based on averages.   For example, suppose one wants to estimate decay centrality with $T=3$, but only using information about first and second neighborhoods.
An approximation (and taking $p_{ij}=p$ for all $ij$) is:
\[
p d_i+ p^2 n_i^2  +  p^3 (n_i^2  (\overline{d}-1)),
\]
where $\overline{d}$ is the average degree in the network. Thus, $n_i^3$ is approximated by  $(n_i^2  (\overline{d}-1))$, which calculates how many nodes are at distance 2 from $i$
and then presuming that each of them has an additional $\overline{d}-1$ friends.\footnote{If one has an estimate for clustering, then one could further refine
the estimate by adjusting this to account for the fact that some of those node's friends may already be in $i$'s second neighborhood.    So if average clustering is $\overline{c}$,  then  the estimation of nodes at distance 3 would be
$ p^2 n_i^2  (\overline{d}-1)  -  2 \overline{c} \frac{n_i^2 (n_i^2-1)}{2}$, where the last expression accounts for the fact that if two nodes at distance
two are connected to each other, then we should lower the count of further friends that we attribute to each of them by one (hence the factor 2).
One could make further adjustments to account for the number of friends that nodes at distance 2 have that are at distance 1:  that is if there are several paths of length 2 from $i$ to some $j$, then several of $j$'s friends have already been counted.
}$^,$\footnote{An even better approximation is to adjust the $\overline{d}$ to be the average degree of neighbors in the network, which is generally higher than the average degree (e.g., see \cite{jackson2016}).  An approximation is $E[d^2]/\overline{d}$.}

Similar approximations can be done for diffusion centrality.\footnote{In that case, clustering and double counting is less of an issue, since all walks are counted,
and so the approximations are even easier.}

\subsubsection{Directed Networks and Information Capital}

For undirected networks, the above definitions are appropriate for applications that involve either sending and receiving information.
With directed networks, however, there is a clear distinction between sending and receiving.    In that case, the definitions above are the appropriate ones for {\sl sending} information.
However, for {\sl receiving} information, one needs to take care to account for paths from other nodes $j$ to reach $i$, rather than the other way around.
So, the appropriate measures become
\[
Dec_i^{Rec}(\mathbf{g},p,T, \mathbf{v}) =  \sum_{\ell=1}^T  \sum_{j: i \in N^\ell_j (\mathbf{g})}
 p^{\ell} v_{ji}
\ \ \ \ \ \ \
{\rm and} \ \ \ \ \ \ \
Diff_i^{Rec}(\mathbf{p},T, \mathbf{v}) = \sum_{\ell=1}^T  \sum_j
[\mathbf{p}^\ell]_{ji}v_{ji}.
\]
Clearly, if $\mathbf{g}$ and $\mathbf{v}$ are symmetric, then these are the same measures as before.

\subsubsection{Which Networks }

In the definitions above (and below) I take the network as a given, but of course people are involved in many different types of relationships and only some of them might be relevant with respect to a given form of social capital.  Information flows are perhaps the most general in that we communicate and gossip with people who we see in all sorts of different roles and relationships, whether they be co-workers, neighbors, close friends, or family.   In \cite{banerjeecdj2013} we used a union of twelve different types of relationships to predict information flows - as it was clear that all those forms of contact also involved some exchange of information that could be completely unrelated to the reason for the contact.  Of course, network measures can be weighted and directed to adjust for the frequency of communication, as discussed above.  Which data are appropriate to account for information flow becomes a largely empirical question and here I provide definitions presuming that a communication network has been properly estimated.
How to gather appropriate data is a subject beyond the current paper, and is perhaps easiest when it comes to communication, and becomes more challenging with respect to some other forms of social capital.

\subsection{The Godfather Index:  Brokerage, Coordination, and Leadership Capital}

Information capital is important as it embodies our ability to be well-informed and to inform others.
What information capital misses is the uniqueness of our position in doing so.   Can people get the same information via others?
Would they lose out if a given person was removed from the network?

The criticality of a given individual in coordinating and connecting others is often a necessary source of power.
For instance, it plays a key role in understanding the Medici's rise to power in fifteenth century Florence \cite{padgetta1993,jackson2008,jackson2018}.
It is part of the idea behind Burt's \citeyearpar{burt1992} well-known concept of structural holes.

Here it is useful to distinguish two more forms of social capital even though they will both be measured by a common index.

One concept is {\sl brokerage capital}  being in a position to serve as an intermediary between others who need to interact or transact.
An individual can play a critical role in enabling a transaction to occur.  This is perhaps best embodied in Mario Puzo's, and Francis Ford Coppola's, fictional character of the
Godfather.   He never performed favors directly, but who was instrumental in connecting other individuals.  His key position meant that he was not only able to connect others, but was also able to collect favors himself.     By granting favors to another he was able to later call on them for help, either on behalf of someone else or on behalf of himself.

A different but related concept is that of {\bf coordination capital}, which might also be termed {\sl leadership capital}:  being situated as a `friend-in-common' to others who cannot coordinate their actions directly, and thus being in a position to coordinate others' behaviors.\footnote{This is a very narrow use of the term leadership, as it stems entirely from a person's positional ability to coordinate the activity of others, but does not involve other personal characteristics or other factors that determines whether people will actually pay attention to the individual.  }

Being in position as a friend in common of people who are not directly connected is particularly important when people need to act collectively or coordinate their actions.   One example was the Medici, who occupied a key position among the main families that comprised the oligarchy in Florence in the fifteenth century.
There was a critical point in time at which the Medici were able to coordinate a number of families to provide armed men and act politically, while their key rivals the Albizzi and Strozzi, were not able to coordinate other families to counter the Medici.  The Medici's position in the network was clearly different from other families, in particular in terms of serving as a connector of other families.\footnote{See the discussion in \cite{kent1978,padgetta1993,jackson2008,jackson2018}.}
More generally, being in a position to bring others together can result in combining them in productive activities when they would not have otherwise come together.\footnote{ For example, see the discussion in  \cite{brassk1999} of the importance of network position and coordination in leadership.
Being able to coordinate others has also been discussed in defining power (e.g., see \cite{pfeffer1981,pfeffer1992}), as well as how unequal allocations can be (\cite*{ketsetal2011}).}

A standard centrality measure that comes to mind for measuring the sorts of capital that I have called brokerage capital, as well as coordination and leadership capital, is via the betweenness centrality measure due to
\cite{freeman1977}.
This counts the fraction of all shortest paths between other pairs of nodes that pass through through the given node:
\[
Bet_i(\mathbf{g}) = \frac{2}{(n-1)(n-2)} \sum_{(j,k): j\neq i \neq k \neq j
} \frac{\nu_\mathbf{g}(i:j,k)}{\nu_\mathbf{g}(j,k)}.
\]

There are a few reasons why betweenness centrality is not well suited for capturing either brokerage capital or coordination capital.
One is the practical problem that it becomes cumbersome to compute betweenness centrality in large networks.\footnote{For a discussion of truncated betweenness centrality measures to address this concern, see \cite{ercsey2012}.}
The more important problems with betweenness centrality as a measure of either brokerage capital or coordination capital are that
(i) it gives equal credit to all intermediaries on a shortest path, regardless of how many intermediaries there are on the path,
and
(ii)
it weights all paths equally, even though nodes at great distances are much less likely to interact via some long chain of intermediaries than nodes who are fairly close to each other.\footnote{One way to
account for these issues is weight paths by their length.
For instance, for some {\sl decreasing} function $f(\ell)$,
$
\sum_{(j,k): j\neq i \neq k \neq j
}  f(\ell_{\mathbf{g}}(j,k)  )\frac{\nu_{\mathbf{g}}(i:j,k)}{\nu_{\mathbf{g}}(j,k)}.
$
}

An easy way to deal with all of these problems at once is to restrict attention to a node's immediate neighbors.
In many settings, these are by far the most likely pairs to be brokered or coordinated.  Moreover, this results in a measure that is more practical to compute in large networks.

The basic idea is to measure brokerage and coordination capital by the {\sl number} of pairs of a person's friends who are not friends with each other.

I refer to this measure as the `Godfather Index':
\footnote{This differs from the sorts of measures that one finds in some network software, e.g. UCINET, which calculate measures based on group assignments and directed interactions. (For more discussion behind those measures see
\cite{gouldf1989}.)   Those are actually closer in spirit, but still different from, what is proposed to measure bridging capital below. }
\[
GF_i (\mathbf{g})= \sum_{k>j}  g_{ik} g_{ij} ( 1- g_{kj})  =   |\{k\neq j: g_{ik}=g_{ij}=1, g_{kj}=0\}|.
\]

The godfather index has an inverse relationship to clustering, but weighted by the number of pairs of a node's neighbors.
\[
GF_i(\mathbf{g}) =  (1-clust_i(\mathbf{g})) d_i(\mathbf{g})(d_i(\mathbf{g})-1)/2 .
\]

In a directed and/or weighted case, the measure becomes:
\[
GF_i (\mathbf{g})= \sum_{k>j: g_{kj}=g_{jk}=0}  g_{ki} g_{ji}
\]
So, this accounts for pairs of people who both can be influenced by $i$ (`listen to $i$', by convention $g_{kj}>0$), but neither of whom influences each other.

The reasoning behind counting numbers of pairs rather than fractions, is that each pair leads to potential rewards for the intermediary.   If one instead normalized by the number of pairs that a node has, then one ends up with a fractional measure of
$ 1-clust_i(\mathbf{g})$,
which has very different interpretations and applications.

This inverse relationship between the godfather index and clustering provides an interesting contrast.   It relates to
the contrast between being a key connector as in Burt's \citeyearpar{burt2000,burt2005} concept of structural holes, compared to having `closed' relationships as in Coleman's \citeyearpar{coleman1988} concept which is often associated with clustering.
The ability of an individual to serve as a broker, or to coordinate the activities of others, is enhanced by a lack of connection between those others.
That is what is picked up with the Godfather index.     The ability of a local group to sanction individuals depends more on having closure, as discussed below in the context of favor capital.

\subsubsection{Variations on the Godfather Index}

There are variations on the Godfather Index that are computationally more intensive, but are of interest.
One issue is that, the Godfather Index does not account for the fact that a pair of $i$'s neighbors $j$ and $k$ might also have some other friend in common who
can serve as an intermediary.    Being a unique intermediary, rather than one of several, may lead to higher levels of rewards and power.

Checking uniqueness requires having information on second neighborhoods, which presents a higher computational burden in enormous graphs, but one that is still manageable in many applications.
There is then a question of how to make the adjustment for competition between intermediaries.   One would just be to divide by the number of intermediaries, which would correspond to an even chance that any one gets to be the intermediary.
Another would be to account for potential competition between the intermediaries -- which in the extreme (as in Bertrand competition) would give no value
to connectors of $j$ and $k$ if there is more than connector.

So these two alternative measures would be

\[
 |\{k,j\in N_i(\mathbf{g}): \ell(k,j)=2, \nu_{\mathbf{g}}(k,j)=1\}|,
\]
and
\[
\sum_{k,j\in N_i(\mathbf{g}): j\neq  k, \ell(k,j)=2
} \frac{\nu_{\mathbf{g}}(i:j,k)}{\nu_{\mathbf{g}}(j,k)}.
\]
This latter corresponds to the ``ego betweenness centrality'' measure discussed by
\cite{everettb2005}.

\subsubsection{Distinctions Between Brokerage and Coordination Capital}

Given that both brokerage and coordination capital involve the same network measure and involve being in a unique position as a connector between neighbors, one might wonder what the real distinction between them is.
The distinction is in how this position is used and how the measure should be evaluated.

It helps to first discuss some examples of brokerage capital.   As one example, a canonical setting in which brokerage capital would be useful would be as an intermediary for negotiations - for instance a country being an ally of two other countries who are facing a conflict with each other and for which the intermediary can serve as a facilitator of negotiations, and being able to pressure each side to compromise.   Another example would be an agent who puts buyers and sellers in contact with each other - such as an agent for film stars, writers, athletes, etc.\footnote{This brings up a refinement of the measure of brokerage capital in cases where markets are bilateral.   In the first example of countries, it could be that any two countries could need to enter negotiation and so the Godfather index properly captures the potential interactions.   In an example of serving as an agent between buyers and sellers, then in counting the Godfather index, one would want to only count pairs of neighbors that involve one buyer and one seller, but discard pairs that involve two sellers or two buyers.   That is, being a unique connector between two sellers might be useless since they do not need to transact.}
In both of these examples, there is usefulness in being able to bring the two other actors together to some sort of contract or negotiation.   Having longer-term relationships with the parties on both both sides can also help facilitate the agreement between them, both because its success affects the reputation of the intermediary and the future relationships that it has with the two parties.
The other thing that is special about these examples is that usefulness of the position is realized in connecting two other parties at a time - serving as a broker between two others.

When it  comes to coordination capital, the idea of leadership often comes from helping a group of other actors come together all at once and act collectively.  For instance, in the example of Cosimo de Medici, it was his ability to raise a militia from other families who were his friends.   Without his presence, the action would have failed - as there would not have been any natural focal family who could have communicated with the others and coordinated their activities and provided sufficient trust in success to get them to act.    In this case, it is no longer bring pairs together, but instead in bring groups of others together and helping them act collectively.    It still requires measuring how many other actors someone serves as unique connector between.

For both types of capital, the Godfather index is a measure of the extent of the position, but then would be evaluated based on some contextual information about the value of the position.    In the case of brokering, what is the expected value of the deals (or each deal)?   What sorts of rewards might accrue?
In the case of coordination and leadership, which situations might coordinating some given group of agents prove to be useful and then how useful?
For example, it might be that a certain threshold of the index must be reached to be successful - connecting three other actors might not be enough to have a successful rebellion, but connecting four other actors might.    So, in some contexts the value of the capital that is measured by Godfather index might not vary linearly in the index, but be a threshold function or some other nonlinear function of it.

\subsection{Bridging Capital}

The next type of social capital that I wish to distinguish is that of
 {\bf bridging capital}:  having a position as a unique or vital connector between otherwise disparate groups, with an ability to acquire and control the flow of valuable knowledge.

What I am referring to as bridging capital is often confused with the brokerage capital and coordination capital notions since it has some similarities in terms network positions.
Nonetheless it is conceptually distinct and can be measured differently.\footnote{%
Bridging capital is perhaps conceptually the most closely related to Burt's \citeyearpar{burt1992,burt2000,burt2005} notion of structural holes, although his discussions include aspects of all three of bridging, brokerage, and coordination capital.}

Bridging between otherwise disparate groups can be very productive in a variety of ways.   One is that information, techniques, or useful norms from one group can be imported to another.   Serving as a bridge enables a person to do such importing.    Beyond the rewards for bringing valuable knowledge or customs to a group, a person might also benefit in other ways.    Such new knowledge can also result in new synergies when put together with existing knowledge of another group (e.g., see \cite*{akcigitetal2016}). \footnote{See also \cite{anderson2017} for an account of how particular combinations of skills and knowledge can be more valuable than separate skills or superficial knowledge of many skills.}

Bridging capital differs from brokerage capital as it is based on being a key conduit of useful knowledge between groups rather than intermediating transactions or coordinating the actions of others.   It is more about controlling and making use of unique combinations of information.
 Bridging capital is also different from information capital, as it is not just an ability to broadcast or receive, but is more dependent
 upon the uniqueness of a person as a conduit.  

 These distinctions become easier to see when we provide a measure for bridging capital.

This is the most difficult measure of social capital to calculate in a network.   One needs a measure that identifies different groups of nodes in a network that are relatively separated and then identify key connectors between those groups and adjust for how exclusive a given individual is in terms of position.   This depends on how big the groups are, and how many connections exist between the groups.  With competing connectors,  any given individual is less necessary in terms of increasing productivity and importing and exporting knowledge and customs and so they have less capital.
Again, a temptation is to use some variation on betweenness centrality, but it fails to capture the things I just mentioned.

Here is a suggested method for calculating how critical a given link $ij$ is as a bridge:  a unique conduit or connector.

Let $\mathbf{p}-ij$ indicate the matrix $\mathbf{p}$ with entry $ij$ set to 0.

 The criticality or importance as a connector of link $ij$ is
 \[
 Crit_{ij}(\mathbf{p},T,\mathbf{v}) = \sum_{i'j'} v_{i'j'} (EInf(\mathbf{p},T) - EInf(\mathbf{p}-ij,T))_{i'j'}.
 \]
 So, we compare how much information flows in the society with and without the relationship $ij$ present.  If no relationship between $ij$
 exits in $\mathbf{p}$, then $Crit_{ij}=0$.

 Clearly, one can use other measures, such as $PInf$ instead of $EInf$, depending upon which is the most appropriate in a given context for capturing the
 importance of information.

 Then the bridging capital of a node $i$ is simply
 \[
 Brid_i(\mathbf{p},T) = \sum_j Crit_{ij}(\mathbf{p},T)   = \sum_j   \sum_{i'j'} v_{i'j'} \left(\sum_{t=1}^T \mathbf{p}^t - (\mathbf{p}-ij)^t\right)_{i'j'}.
 \]
 So this accounts for all the possible bridging of information that node $i$ does in the network.

 Although this is obviously more complex than calculating information capital, it just involves repetitions of such calculations, and so in large networks one can work with approximations of the information capital to define the criticality of each link, and then there is a summation across links.  In huge networks, this might require very simple approximations of information capital (for instance, limited to $T=2$) in order to be practical.

 This measure captures not just the role that $i$ plays in relaying information, but how that changes when $i$ is removed from the network.
 This captures how exclusive $i$ is, since if many other paths exist in a network, then a given $i$'s relative level of bridging capital will be low.
 \footnote{This is a much more direct measure than something like the
effective size of \cite{burt1992}, which does not capture exclusivity.  Effective size computes a node's degree and then subtracts the average degree of
that node's friends.   Having many friends who are not well connected themselves may be correlated with bridging capital, but is still less directly tied
to bridging capital than the measure above.  Similarly, effective size is less directly related to brokerage or coordination capital than the Godfather Index.
}

The above measure of bridging is an example of a technique that can be varied to provide a whole family of measures.
It consists of two steps.
In the first step one assigns some value to each link as a connector,
and then in the second step each individual gets the value of the sum of his or her links.
 This is an approach to calculating centrality described by \cite{everettv2016}.
The definition for Bridging capital above uses a different way of assessing a link's value as a connector that in \cite{everettv2016}, but is in the same spirit.

\subsection{Favor Capital and Support}

Another form of social capital is {\bf favor Capital}:
the ability of an individual to exchange favors and safely transact with others
through a combination of network position and repeated interaction.

In many instances, we might see the connections but not have good measures of the repetition of favor exchange or other transactions.
We can look to the network structure to see how many friends an individual has, and how well those friendships are supported.

The basic theory behind this is straightforward:   consider a friendship between two people $i$ and $j$.
The more frequently they reciprocate in exchanging favors, the more valuable the friendship becomes and the more incentives they have to
continue to exchange favors.
While such repeated interaction can work well with relatively small favors that can be exchanged with high frequency, it is generally harder to maintain the exchange of large favors that can be very costly to provide and are not likely to be reciprocated for some time.

Such exchange is then facilitated by having friends in common.   If $j$ is asked to provide a favor for $i$ but then does not provide it, it is not only $i$ that can refuse to reciprocate in the future, but also other friends that the two have in common.   This provides greater incentives for $j$ to deliver favors when asked.
The theory backing this is provided in \citet*{jacksonrt2012}.\footnote{For further discussion of how incentives for reciprocation and cooperation are dependent upon network structure,  see \cite*{raubw1990,blochgr2007,blochgr2008,karlanmrs2009,lipperts2011,alim2012,alim2013}.}

We say that a relationship between $i$ and $j$ is supported if they have at least one friend in common.

Favor capital can be measured via support, counting the number of friends that a person has who are supported by a friend in common:
\[
Supp_i = |\{j\in N_i(\mathbf{g}): [\mathbf{g}^2]_{ij}>0\}|.
\]

This is an absolute count of the number of supported relationships, rather than the fraction of supported relationships, as the absolute count is more directly related to the total favor capital that an individual has.
There is also a sense in which supported relationships are more likely to be `strong ties' in the sense of \cite{granovetter1973}, while weaker friends with whom one rarely interacts are less likely to be supported.   Such weaker relationships can be very valuable when it comes to information capital, but are less likely to be valuable as a source of costly favors.

This is a very local measure, and hence one that is not computationally challenging in large networks - one needs to be able to cross check lists of friends (effectively seeing who is reachable via paths of both length one and two).

Although support shares some common features with Coleman's notion of closure,
which plays a role in his famous discussion of social capital \citep{coleman1988},
it is important to emphasize that support is distinct from closure both in concept and in how it is measured.
Coleman's closure is often associated with clustering - that a given node $i$ can be sanctioned if
more of $i$'s friends are friends with each other and can thus coordinate their behaviors towards $i$.
Support is a measure that applies to a {\sl relationship} rather than an individual.  The support measure then counts
how many supported relationships an individual has.

\begin{figure}[h!]
\begin{center}
\includegraphics[height=2in]{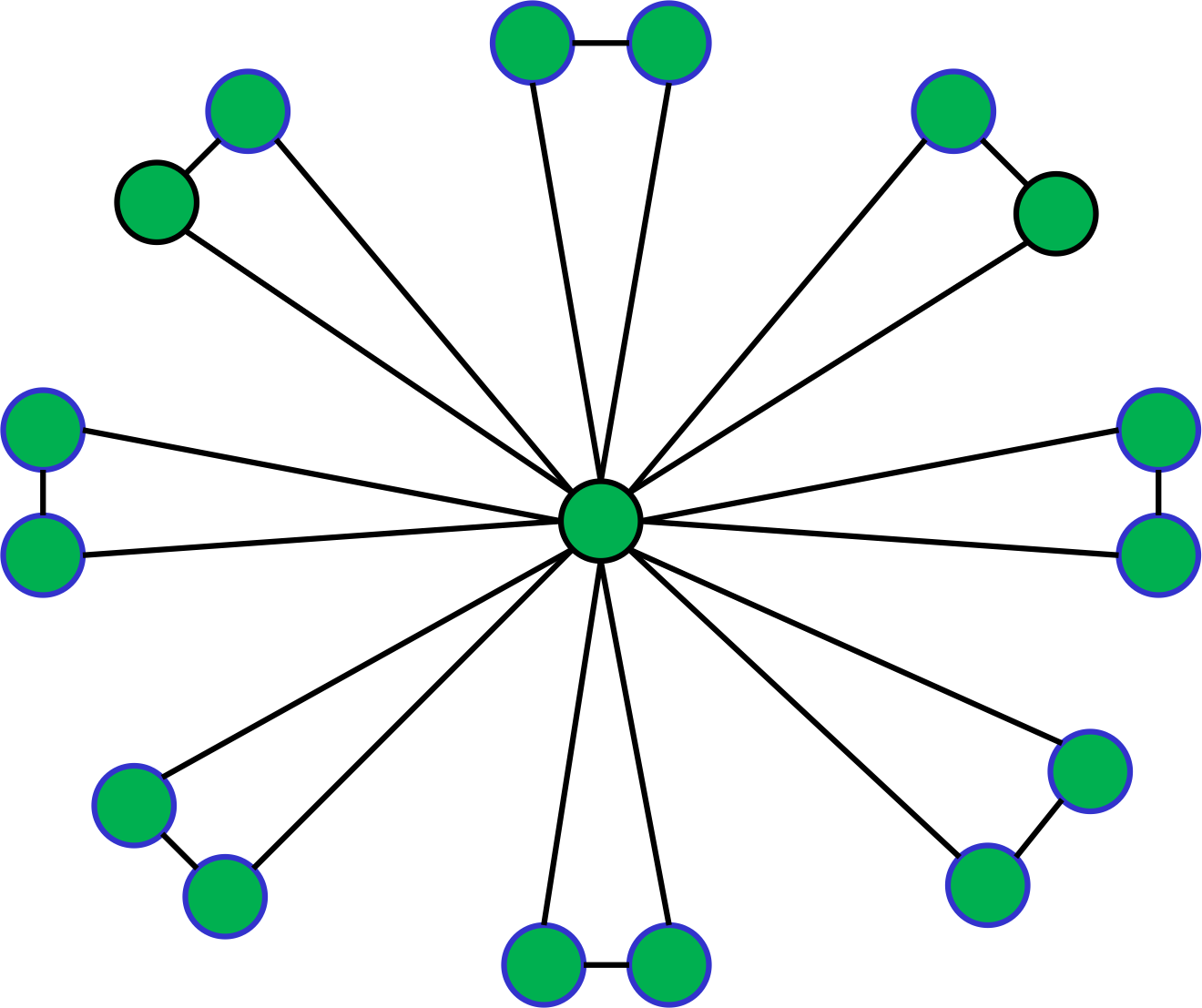}
\caption{\label{support}
\small{The difference between support and clustering.   In this network, every one of the center's nodes are supported, while the clustering coefficient is only 1/15.  In the limit, if we kept adding such triangles centered at the middle node, clustering would tend to 0, while every one of the center node's relationships would be supported.
}}
\end{center}
\end{figure}

It is easiest to see the distinction between support and clustering in Figure \ref{support}.
In that figure, every one of the central node's relationships is supported -- the central node has a friend in common with every one of his or her friends.
Yet, the clustering is only 1/15 -- most of the friends of the central node are divided into distinct subgroups with little overlap.
The central node has a high level of support, but a low level of clustering.

Favor capital should not be confused with brokerage capital, even though they both involve favors.
Brokerage capital comes from being a key intermediary, while that is not at all required of favor capital.   For instance, in a setting in which an individual has many friends and they are all connected to each other, then the number of supported relationships would be very high and favor capital would be high and yet brokerage capital would be zero.
In contrast, an individual at the center of a large star network in which none of the relationships are supported would have a high Godfather index, and so brokerage capital would be high and yet favor capital would be low.
The distinction is that brokerage capital comes from being an intermediary and delivering indirect favors and advice, while favor capital is about being counted on to personally reciprocate in providing favors and advice.

The measure of support is also a particular one.  One can enhance it in several directions.   First, relationships that are more heavily supported - supported by many friends in common may be stronger ones that are more reliable.   For instance, one can also count the level of support by tracking the number of friends in common, giving a higher support score for relationships that have more friends in common - so there would be a weight multiplying each supported relationship that would score how highly supported it is.   Second,  it could be that there are interactions between relationships.    If you have a supported relationship with someone who has hundreds of other supported relationships, that may count for either less than having a supported relationship with someone who only has a few other supported relationships.   For instance, someone having many other options may require fewer favors from you and hence may be less reliable in reciprocating.    Third, there is also an obvious heterogeneity in favor types:  a friend who is a doctor could be counted on for medical advice, but might not be very good at providing financial advice.   Different types of favors may have different values, and this abstracts from that heterogeneity.   Attaching values to nodes, or directed pairs of nodes (how much $i$ values favors from $j$), would then capture this.

Support captures some core facets of favor capital but obviously not all of them, as there are various reasons that an individual can rely upon others for favors besides having supported relationships, including having some organizational or social status or other reputational concerns.  To the extent that such features are important and measurable, they should clearly also be included in measuring favor capital.  Support is meant to be a relatively easy and direct network measure.

\subsection{Reputation Capital}

A reputation, in the form of the beliefs of others for how a given individual or organization will perform in producing a good or service, can be a very valuable
form of capital (e.g., see \cite*{mailaths2006,klewesw2009})   Many production processes involve multiple producers and reliance on others.   The value of the process can depend on how well those individuals perform their respective jobs.   In settings where there is uncertainty about how well someone might perform, having a `good' reputation for reliable and consistently high performance can enable production to occur when it might not otherwise take place.   It can help in all aspects of production, from securing the funding and collaboration to get a process started, to the end marketing and distribution of a product.
It can also be instrumental at an individual level, in getting a person hired or included in some productive activity.  It can change the way that an organization or person is treated and the way that others interact with it.

It is clear that reputation should be considered as a form of capital:  it is a valuable input into production, is a stock that can be acquired over time and used, or converted into something that can be used, for productive purposes.    It is also clear that it is a social form of capital:  it is not physical nor embodied in the individual(s) who possess it - a reputation depends on the beliefs and perceptions of other people in the society.

The fact that reputations are dependent upon beliefs and expectations of individuals makes them particularly hard to measure, since beliefs and expectations
are hard to measure under the best of circumstances.   There is no obvious and easy network or structural measure of reputation.   One can survey people, or otherwise measure reputation indirectly, but it is not something that
is easily captured from data solely on interactions.   Measuring reputation is its own literature that takes us beyond the current paper,\footnote{For a few starting references on measuring reputation see \cite{list2006,berens2004,wartick2002}.} and so I simply note that reputation is another form of social capital.

\section{Community Level Social Capital}\label{community}

The various forms of social capital defined above can be aggregated across individuals to understand how much capital is possessed by
an organization or group of people.    For instance, one can define community-level information capital,  community-level favor capital, community-level bridging capital, etc., by aggregating or averaging individual measures.
However, when aggregating, there can sometimes be better ways to track the community total than by simply summing across the individuals.  Simple summing might involve some double counting.   For instance, when measuring communication centrality from a set of nodes, one can directly calculate the number of other nodes outside of a community that end up informed if all members of a community start with information and then randomly spreading it.  This results in a different number than summing across the people
   in the community, and avoids double counting.

As an example of the importance of community levels of social capital, a measure of how many links go outward from a community  has been found by \cite*{baileyetal2016} to correlate with a number of important outcomes, such as economic mobility, education rates, crime, and other behaviors.

However, beyond being careful in providing community level measures of the various forms of social capital defined above, there is also a very different concept of community capital that differs from
an aggregate measure,   but instead relates to the internal functioning of a community.

\subsection{Community Capital}

This last form of social capital is one that has been studied from various aspects, dating to at least   \cite{banfield1958}, and perhaps closest in spirit to what was originally meant by Lydia Hanifan \citeyearpar{hanifan1916,hanifan1920}.

I will define {\bf community capital} to be the ability to sustain cooperative (aggregate social-welfare-maximizing) behavior in transacting, the running of institutions, the provision of public goods, the handling of commons and externalities, and/or collective action, within a community.

This is perhaps the most difficult form of social capital to measure as it is not embodied in a social structure and applies to a group of people who are typically not isolated from interactions with non-group members.  It is partly embodied in the norms and culture of a community.
It is also not something that an individual can ``possess'', but is a feature of a society.
It comes from a combination of things, including a selection of a norm or coordination of behavior within a society, as well as a robustness of that behavior so that it is sustainable over time and in response to inevitable shocks and perturbations.\footnote{The fact that individuals in a community are more familiar with others in the community can also be productive, as it can help them predict how others will act and hence to coordinate their behaviors which can be productive (e.g., see \cite{jacksonx2017}).}
It may require certain forms of organization or institutions as well.

Researchers have tended to work with some measure of  ``trust'' to measure what I am referring to as community capital, and trust has been found to correlate with various outcomes
and measures of a society's success
(e.g.,  \cite*{banfield1958,knackk1997,putnam2000,hamptonw2003,tabellini2010,nannicinietal2013}).

Despite the challenges in measuring this form of community capital, it can have large welfare implications and so is important to better-understand.
The complexity of the interaction of all the factors that affect how well a society functions makes this an ongoing area of research.
Community capital will thus will require further parsing and study, and I leave that for future research.

\section{Concluding Remarks}

The typology here is meant to help provide a basic listing of, and distinctions between more fundamental forms of social capital that have all been discussed, and often lumped together, under the broader umbrella of social capital.
Providing explicit measures helps tie things down and aids in applying the terms to data.  The cost is that any single explicit measure inevitably misses some aspects of the concept it is meant to measure.  However, when operationalizing the concepts, one must make choices, and so I have done that here with the hopes of providing some useful foundations.

In moving to applications, measuring any form of capital is challenging. How does one capture the human capital that an individual possesses?   Counting education and work experience provide proxies, but are far from exhaustive measures of a person's knowledge and skills.   How does one properly value all of the physical capital in the form of plant and machinery that a manufacturer possesses?  It is not as straightforward as one imagines, as their value depends on the wide variety of surrounding circumstances that dictate how they could be most effectively used.

Measuring various forms of social capital is especially difficult as they are dependent upon relationships between people, which are often intangible and only indirectly observed.
The measures of the forms of social capital defined above are thus imperfect and leave many contextual details open.   They are meant to provide anchors for further elaboration and extension.
More generally, this typology is not meant to be the last word on the subject, but to spur the literature forward both in developing additional theory and in applying the concepts.

\bibliographystyle{jpe}
\bibliography{socialCapital}

\end{document}